# On the quantum spin Hall gap of monolayer 1T'-WTe$_2$

*Feipeng Zheng, Chaoyi Cai, Shaofeng Ge, Xuefeng Zhang, Xin Liu, Hong Lu, Yudao Zhang, Jun Qiu, Takashi Taniguchi, Kenji Watanabe, Shuang Jia, Jingshan Qi, Jian-Hao Chen,[*] Dong Sun[†] and Ji Feng[‡]*

E-mail:
[*] chenjianhao@pku.edu.cn;
[†] sundong@pku.edu.cn;
[‡] jfeng11@pku.edu.cn;

F. Zheng, C. Cai, S. Ge, X. Liu, H. Lu, Y. Zhang, J. Qiu, Prof. Dr. Shuang Jia, Prof. Dr. Jianhao Chen, Prof. Dr. Dong Sun, Prof. Ji Feng
International Center for Quantum Materials, School of Physics, Peking University
Beijing 100871, P.R. China
Collaborative Center for Materials Innovation
Beijing 100871, P.R. China

X. Zhang
School of Physics and Engineering, Zhengzhou University
Zhengzhou 450001, P.R. China

Prof. Dr. T. Taniguchi, Prof. Dr. K. Watanabe
High Pressure Group, National Institute for Materials Science,1-1 Namiki, Tsukuba, Ibaraki 305-0044, Japan

Prof. Dr. Jingshan Qi
School of Physics and Electronic Engineering, Jiangsu Normal University
Xuzhou 221116, P. R. China

Quantum spin Hall (QSH) materials are two-dimensional systems exhibiting insulating bulk and helical edge states simultaneously. A QSH insulator processes topologically non-trivial edge states protected by time-reversal symmetry, so that electrons can propagate unscattered. Realization of such topological phases enables promising applications in spintronics, dissipationless transport and quantum computations. Presently, realization of such QSH-based devices is limited to complicated heterostructures. Monolayer 1T'-WTe$_2$ was predicted to be semimetallic QSH materials, though with a negative band gap. The quasi-particle spectrum obtained using hybrid functional approach shows directly that the QSH gap is positive for monolayer 1T'-WTe$_2$. Optical measurement shows a systematic increase in the interband relaxation time with decreasing number of layers, whereas transport measurement reveals Schottcky barrier in ultrathin samples, which is absent for thicker samples. These three independent pieces of evidence indicate that monolayer 1T'-WTe$_2$ is likely a truly 2-dimensional QSH insulator.

The theoretical proposal and experimental discovery of topological insulators [1-3] have provided an invigorating momentum to condensed matter and materials physics. In two dimensions, a topological insulator displays the QSH effect[4-8], in which the symmetry-protected helical edge modes are gapless and can propagate unscattered[9, 10]. Experimentally, edge conductance quantization has been demonstrated in QSH phases created in semiconductor quantum wells, such as HgTe/CdTe and InAs/GaSb systems[8, 11]. Free-standing 2-dimensional QSH insulator is particularly attractive[6, 12, 13]. Realization of such topological phases will enable a multitude of possibilities, owing to the chemical stability, mechanical transferability of truly 2-dimensional materials, and their ability to form vertical junctions in van der Waals heterostructures[14]. Despite of the obvious advantages, the realization of truly 2-dimensional QSH insulator has proved challenging.

Recently, a family of transition metal dichalcogenides (TMDC), monolayer 1T'-MX$_2$ (M = Mo, W, X = S, Se, Te), were predicted to be QSH materials, based on density-functional theory calculations[13]. TMDC's are robust layered compounds that can be mechanically exfoliated, and therefore very promising for achieving stable 2-dimensional QSH phase. However, only monolayer 1T'-MoTe$_2$ and 1T'-WTe$_2$ were predicted to be semimetals with negative QSH gaps among the six compounds. Taking a *k*-dependent chemical potential to separate occupied and unoccupied manifolds, the same authors identified the Z$_2$ invariant[1-3, 7] of the "valence bands" to be non-trivial. It is certainly desirable to have a positive QSH gap, which will allow the manifestation of helical edge modes unhybridized with bulk channels. It should be noted that these calculations were performed using the density-functional theory, on which many-body correction is taken into account at the level of G$_0$W$_0$. The fact that density-functional theory generally underestimates band gap calls for further analysis of the quasiparticle spectrum of monolayer 1T'-WTe$_2$. Recently, a related compound, MoTe$_2$ is shown to be metastable in the 1T' structure (natural and stable structure of this compound is 2-H phase), and a band gap is possibly opened up in this material due to strong spin-orbit coupling (SOC)[15]. Consedreing SOC effect is stronger in WTe$_2$ system, monolayer 1T'-WTe$_2$ is expected to have larger band gap and requried further investigation.

In this work, we report three independent pieces of evidence for the existence of positive QSH gap in monolayer 1T'-WTe$_2$, from both theoretical and experimental viewpoints. We employ the hybrid functional method, in which the density-functional theory is supplemented by exact exchange functionals with screening correction, to evaluate the quasiparticle spectrum. Our calculations show that the many-body effects and SOC together give rise to a positive QSH gap of monolayer 1T'-WTe$_2$. Although monolayer sample of 1T'-WTe$_2$ remains evasive experimentally, we perform systematic optical and transport measurements on samples with thickness down to 9 – 11.7 nanometers (nm). We find that the interband carrier relaxation time increases dramatically as the sample becomes thinner, whereas the intraband relaxation time remains largely constant. The transport measurement reveals that in an 11.7-nm-thick sample, the I-V curve becomes patently non-linear below 10 K, from which a Schottky barrier ~ 1.76 meV is extracted. The calculated spectrum suggests that that monolayer 1T'-WTe$_2$ has an interaction-induced positive QSH gap, which are supported by the experimental results.

Density-functional theory calculations were performed using VASP[16], with the generalized-gradient approximation, parameterized by Perdew, Burke, and Ernzerhof (PBE)[17], to investigate the geometry and electronic structure of n-layer 1T'-WTe$_2$ (n = 1-4). The Kohn-Sham single-particle wavefunctions were expanded in the plane wave basis set with a kinetic energy truncation at 400 eV. The zone sums were performed on an 8×16×1 **k**-grid for the calculation of the monolayer and bilayer, and on a 4×8×1 **k**-grid for the trilayer and four-layer, with the 1$^{st}$-order of Methfessel-Paxton type broadening of 0.1 eV. We employed a vacuum region of larger than 15 Å, to ensure sufficient separation between the periodic images of 1T'-WTe$_2$. The standard HSE06 hybrid functional incorporating a screened Coulomb potential[18-20] was also used to determinate the equilibrium crystal structure and electronic band structure of the monolayer. SOC is taken into account in the structural optimization and band structure calculation of the monolayer. The equilibrium crystal structure is relaxed with a conjugated-gradient algorithm, until Hellmann-Feynman forces on each atom are less than 0.02 eV/Å. As van der Waals interaction is likely to be important only in multi-layer structures, corrections for long-range correlations are not included in evaluating the total electronic energy.

We first examine the structures of the monolayer optimized using PBE alone, PBE with hybrid functional (HSE06), with and without SOC. The optimized lattice parameters *a*, *b*, are shown in **Table 1**. It is seen that the lattice becomes slightly compressed when HSE06 is used, compared with the structures relaxed by PBE functional. When SOC is included, the lattice also becomes slightly compressed in *a* direction but stretched in *b* direction. In the HSE06 implementation, the exact exchange potential is mixed with a screened version exchange potential from the density-functional theory. The HSE06 hybrid functional uses an optimized mixing of exchange potentials to closely reproduce piece-wise linear dependence of the energy functional on particle number. Therefore, this method not only yields more accurate band structure and quasiparticle gap, but also works well even when the single-particle Kohn-Sham ground state has a negative band gap[21, 22]. For example, J.Paier et al. applied this method to various materials including narrow band gap semiconductor, insulator as well as metal. They found that the method showed excellent performance on the equilibrium lattice constants and notable improvements of band gaps for the materials with narrow gaps[23]. In particular, it was found that the HSE06 functional gave a very good estimate of band gaps and spin-orbit splitting in a set of 23 semiconductors[24]. Therefore, the band gaps, $E_g$, in this work are computed using HSE06, and are found to be all positive regardless of whether to include the exchange-correlation function used for structure optimization. And in all cases, the computed $E_g$ fall into the range between 100 and 141 meV for the monolayer (see Table 1).

**Table 1.** Lattice parameters of monolayer 1T'-WTe$_2$ obtained by different optimization methods and the corresponding band gaps.

| Method | a [Å] | b [Å] | $E_g$ [meV][a)] |
|---|---|---|---|
| PBE | 6.316 | 3.491 | 100 |
| PBE with SOC | 6.311 | 3.503 | 136 |
| HSE06 | 6.273 | 3.469 | 108 |
| HSE06 with SOC | 6.268 | 3.481 | 141 |

a) Band gaps are computed using HSE06+SOC on as-optimized structures.

In Figure 1b, we show the band structure of monolayer 1T'-WTe$_2$ with and without SOC. The structure is optimized using HSE06 with SOC. It is found that monolayer 1T'-WTe$_2$ is a gapless semimetal without SOC. Only when both many-body interaction that is described effectively by the hybrid-functional and SOC are included, an indirect band of 141 meV emerges in the spectrum. The result shows that the top of valence band is located at the Γ, and the bottom of the conduction band is located at (0, 0.146). Therefore we conclude that the band gap opening in our calculation is a consequence of both many-body interaction and relativistic effects. It must be emphasized that the computed positive gaps are not result of in-plane strain as discussed in the Reference 13, since the structure optimized with PBE functional is essentially identical to the equilibrium structure reported in the Reference 13, and already has 100 meV gap when HSE06+SOC is used to computed the band structure.

We also investigate the thickness dependence of the band structure of few-layer 1T'-WTe$_2$. For the calculations of n-layer 1T'-WTe$_2$ (n = 2-4), the geometric structures were first obtained by truncating bulk 1T'-WTe$_2$ and the structure of four-layer 1T'-WTe$_2$ was displayed in Figure 1c. The structures of n-layer (n = 1-3) 1T'-WTe$_2$ can be obtained by removing redundant layers. Van der Waals interactions between the adjacent layers were taken into account by using PBE with optB86[25, 26] as exchange functional during the structure optimization of the few-layer 1T'-WTe$_2$. OptB86 is shown to be a reliable exchange functional for the Van der Waals correlation in 1T'-WTe$_2$ system, as the optimized bulk structure with PBE+optB86 agrees well with experimental structure, especially the interlayer spacing is 2.888 Å vs 2.884 Å from experiment. HSE06+SOC is used to compute the bandstructures of n-layer 1T'-WTe$_2$ (n = 2-4) and the results are shown in Figures 1d, 1e and 1f respectively. The corresponding band gaps of the bilayer, trilayer and four-layer are +4, -29, -63 meV as displayed in Figure 1g. It seems that the decrease of the layer number, the increase of the band gap, which agree nicely with our experimental observations, to be presented shortly.

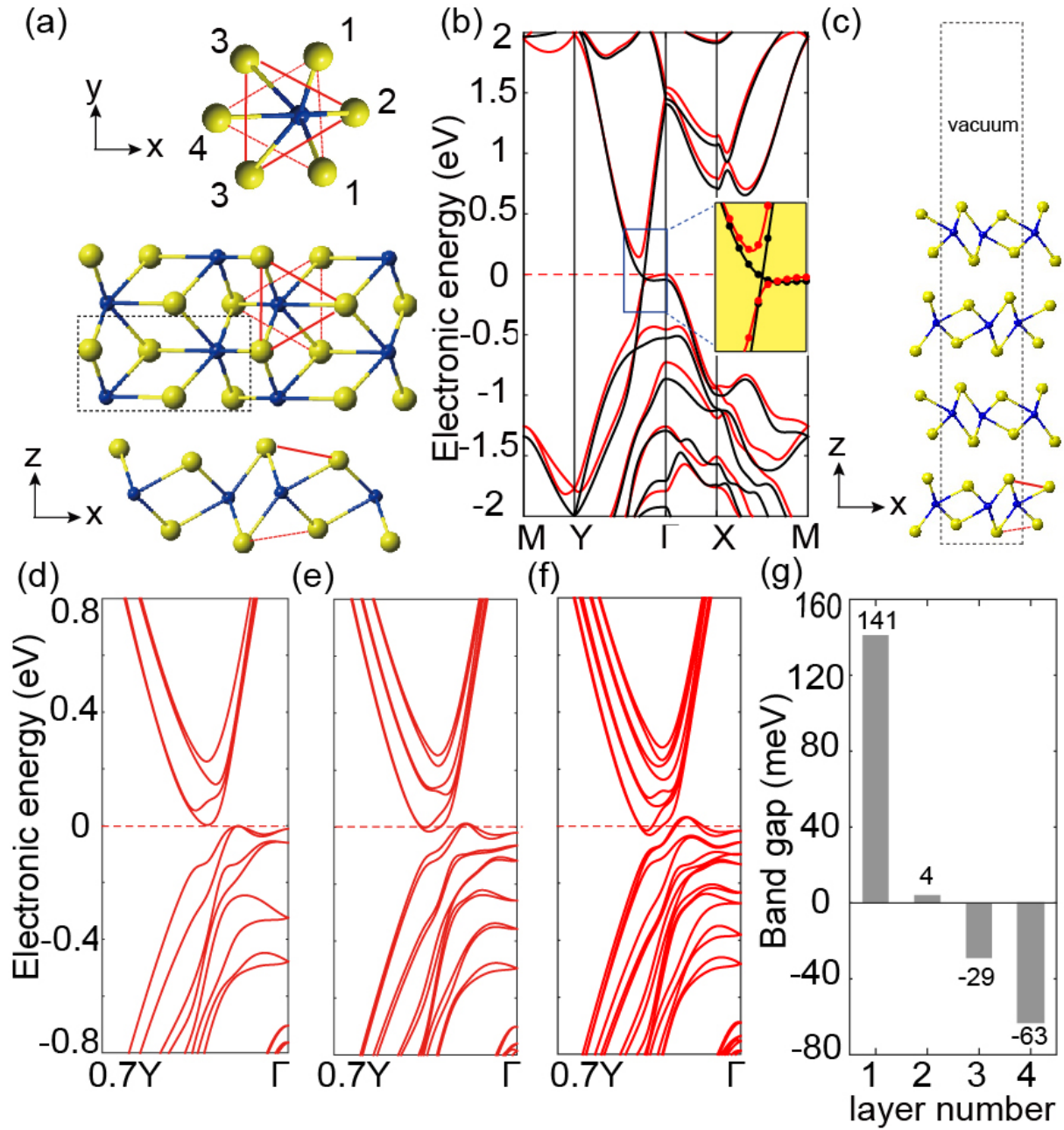

**Figure 1.** Geometric and electronic band structure of n-layer 1T'-WTe$_2$ (n = 1-4). (a) Three views of crystal structure of monolayer 1T'-WTe$_2$ obtained by structure optimization using HSE06+SOC. Yellow balls stand for Te, and blue for W. In the top panel, the distorted octahedral coordination of W by six Te is shown, where one pair of rotated triangles formed by Te are shown with red lines. The W-Te$_i$ bond lengths are, respectively, 2.81, 2.81, 2.70, 2.70 Å for $i$ = 1-4. Black rectangle indicates a unit cell. (b) Band structure of monolayer 1'-WTe$_2$ computed using HSE06, with (red line) and without (black line) SOC. (c) Slab model of four-layer 1T'-WTe$_2$ used in the calculations, consisted of the crystal and an about 15 Å

vacuum layer inserted. Black rectangle indicates a unit cell. The model of n-layer (n = 1-3) can be obtained by removing redundant layers. (d, e and f) Band structure of bilayer (d), trilayer (e) and four-layer (f) 1T'-WTe$_2$ computed using HSE06+SOC. The scales of y axis in (d), (e) and (f) are equal. (g) Computed band gaps of the n-layer 1T'-WTe$_2$ (n = 1-4) using HSE06+SOC, extracted from (b), (d), (e) and (f) correspondingly.

While the many-body effect opens a quasi-particle gap in monolayer 1T'-WTe$_2$, its topological character remains intact. As monolayer 1T'-WTe$_2$ processes inversion symmetry, the topology of the occupied bands can be straightforwardly evaluated through the parity of valence bands at the four time-reversal invariant points, (0, 0), (0, π), (π, 0) and (π, π)[2]. We find that parity of valence bands remain unchanged with and without the hybrid functional correction. The two-dimensional $Z_2$ invariant $v_0 = 1$ is obtained from the product of the parity eigenvalues at the time-reversal invariant points. The robustness of the $Z_2$ is expected, since band inversion occurs at Γ, where the gap is already 141 meV. The interaction effect introduced in the hybrid functional calculations is far less than this energy scale. These results indicate that monolayer 1T'-WTe$_2$ is a genuine QSH insulator with a positive gap. This makes monolayer 1T'-WTe$_2$ all the more attractive, as there is 141 meV window in which the symmetry-protected edges states probed and utilized, without involvement of the bulk states in the transport. We also compute $v_0$ of bilayer, trilayer and four-layer 1T'-WTe$_2$ and the results are 0, 1 and 0 respectively.

To further confirm the topological nature of monolayer 1T'-WTe$_2$, we compute the band structure of a nanoribbon to examine the edge modes. From the HSE06 calculations, we obtained the Wannier functions corresponding to the *p* orbital of Te and d orbitals of W using the direct projection formalism[27]. The band structure for the nanoribbon is obtained using the tight-binding Hamiltonian based the Wannier functions. As shown in Figure 2a, the ribbon is periodic in the *y*-direction but truncated in the *x*-direction with Te-termination, and 20 unit cells are included along *x*-direction in the tight-binding Hamiltonian to ensure sufficient separation between edges. The resultant band structure of the ribbon is displayed in Figure 2b. It is seen that at the time-reversal invariant momenta, Γ and Y, Kramer's degeneracy is present, and that the edge modes (red bands) cross the Fermi three times between Γ and Y, at *A*, *B* and *C* as indicated in Figure 2b. The wavefunction amplitudes at *A*, *B* and *C* are shown in Figures 2c and 2d, reassuring that these modes are well localized near the edge. Modes *B* and *C* are more localized than mode *A*, which is reasonable since the latter is closer to bulk bands. The presence of three bands at the Fermi level between Γ and Y attests to the helical nature of the edge modes. That three modes lie right inside the bulk quasiparticle gap confirms that the bulk is a QSH insulator, of which the transport of the helical edge modes can be probed without intereference from bulk states.

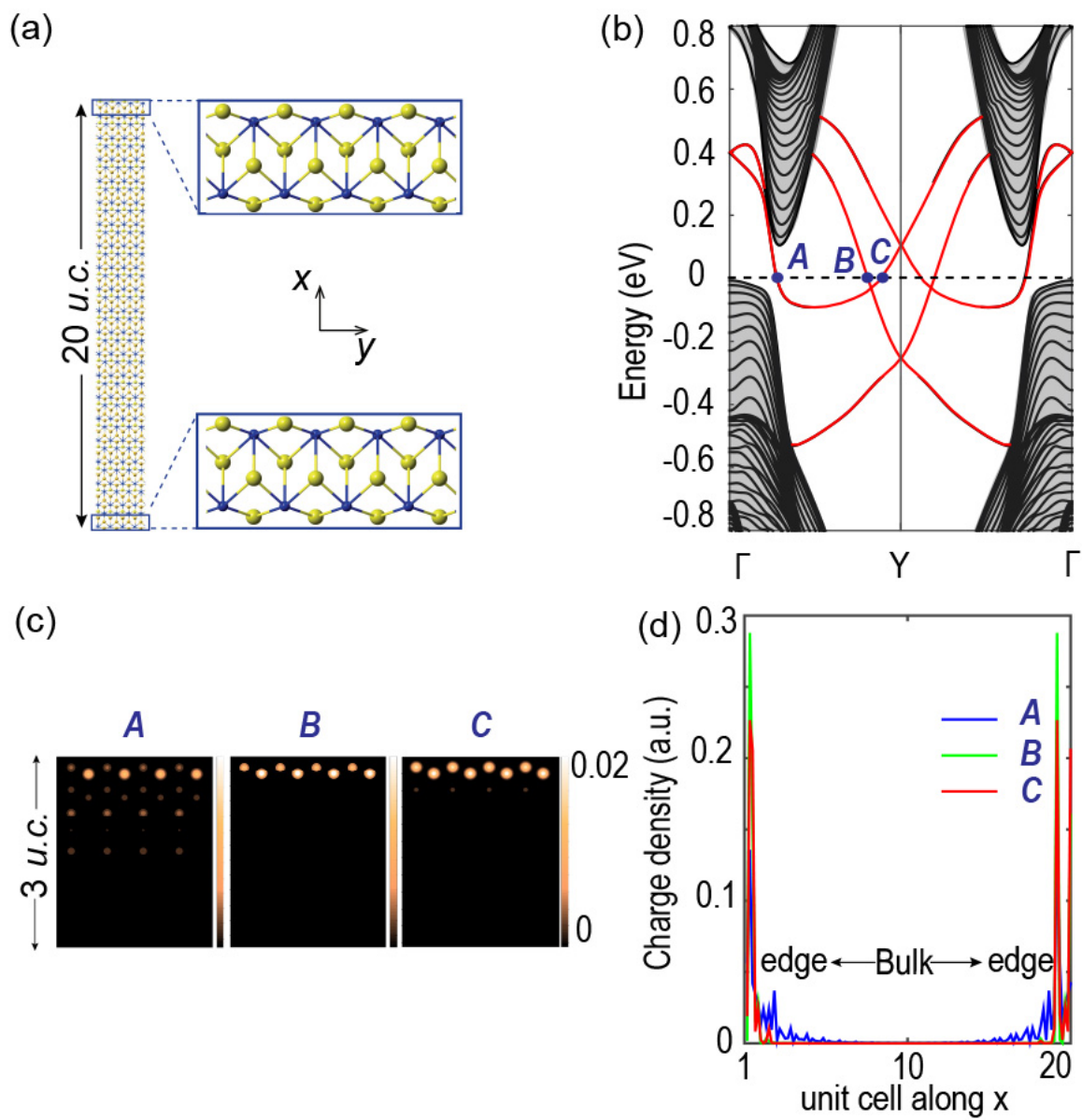

**Figure 2.** Geometry and edge states of monolayer 1T'-WTe$_2$ nanoribbon. (a) The geometry structure of the ribbon, where the blow-up views show the Te-terminated edges. (b) Band structure of the ribbon. Bulk bands are indicated with gray shading, and red bands are localized at the edges. (c) The wavefunciton amplitudes of three edges modes at A, B and C in (b). (d) The wavefunction amplitude integrated along the *x*-direction.

Monolayer 1T'-WTe$_2$ is hard to obtain experimentally, potentially due to stronger interlayer adhesion compared with other TMDC, e.g., MoS$_2$. Nonetheless, we are able to thin 1T'-WTe$_2$ to about 9 nm (~12 monolayers) to 80 nm (~110 monolayers). By performing optical and transport measurements, compelling evidence for possible band gap opening in few-layer 1T'-WTe$_2$ can be obtained, as we report below.

The bulk 1T'-WTe$_2$ sample used in this experiment was characterized by X-ray diffraction (XRD) and the result was displayed in Figure 3a. All major peaks agree nicely with the previous publication[28], which confirms that the bulk WTe$_2$ crystallizes in space group Pmn2$_1$ (1T' structure). Furthermore, Raman scattering measurements were carried out to characterize the ultra-thin WTe$_2$ flakes after they are exfoliated from bulk crystals and fabricated into nano-devices (the data shown was from a 10.3-nm-thick flake). The bulk crystals are characterized with Raman spectroscopy as well. All of our Raman scattering measurements were carried out on freshly cleaved crystal surface perpendicular to **c** axis using the $\lambda$ = 514 nm line of an Ar laser for excitation. The corresponding results were displayed in Figure 3b. Four prominent peaks marked by $A_2^4$, $A_1^8$, $A_1^5$ and $A_1^2$ are observed, which agree well with previous results[29, 30], indicating that our samples used in the transport measurements are high quality 1T'-WTe$_2$ crystals.

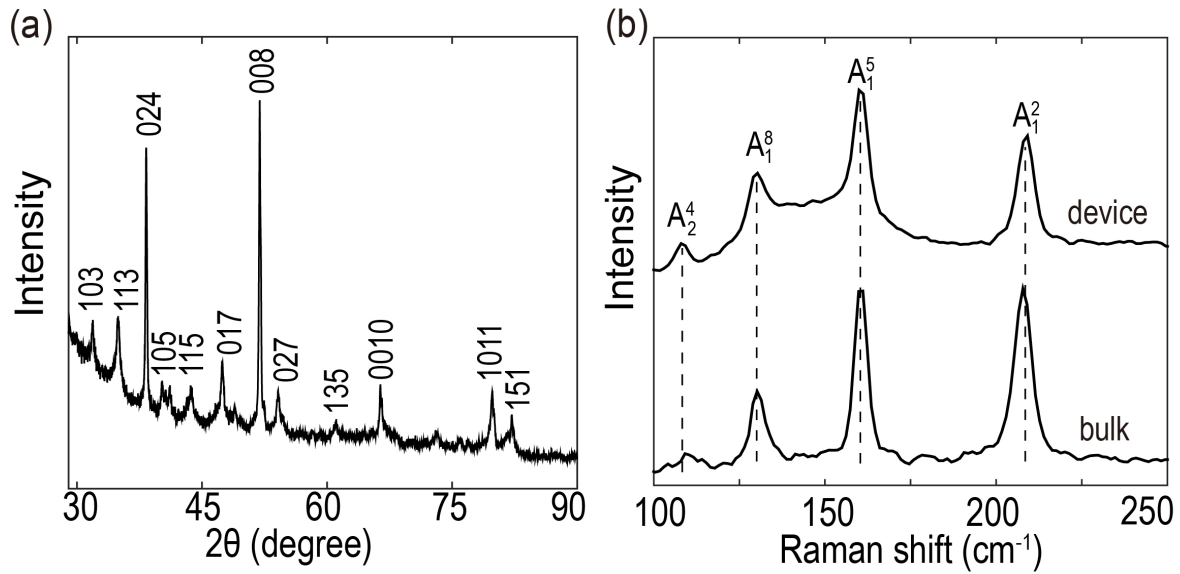

**Figure 3.** XRD and Raman data of 1T'-WTe$_2$ samples. (a) XRD data of bulk 1T'-WTe$_2$ used in the pump-probe measurements. (b) Raman spectra of a bulk 1T'-WTe$_2$ sample and a 1T'-WTe$_2$ nano-device (after the device fabrication process) with thickness of about 10.3 nm. The four visible peaks at about 109, 130, 160, 208 cm$^{-1}$ were marked as $A_2^4$, $A_1^8$, $A_1^5$ and $A_1^2$ respectively. The nano-device and the bulk sample used in Raman measurements were fabricated using the same technology as those used in the transport measurements.

In the first experiment, we employ an ultrafast pump-probe apparatus to examine the photocarrier relaxation using 800 nm photon as pump and 1940 nm photon as probe. The setup of the optical measurement is shown schematically in Figure 4a. The sample used in this experiment is obtained through mechanical exfoliation from bulk 1T'-WTe$_2$ synthesized through vapor transport method[31]. Figures 4b and 4c show the optical and atomic force microscopic images of a typical exfoliated sample with 20-nm thickness. To clarify the rationale of this measurement in regard to the possible gap opening in ultra-thin 1T'-WTe$_2$, we show schematically the relaxation pathways of photo-excited carriers in monolayer and bulk 1T'-WTe$_2$ after pump excitation in Figures 4d and 4e respectively. In the transient reflection measurement, the measurement of $\Delta R$ as function of pump-probe delay $t$ provides direct information about the relaxation dynamics of pump-excited carriers. The relaxation can be divided into three characteristic processes, intraband photocarrier cooling, electron-hole recombination and eventual equilibration with the lattice, respectively, associated with three relaxation times $\tau_1, \tau_2$ and $\tau_3$. In particular, it is well established that as band gap changes, the carrier relaxation displays characteristic evolution for metal[32] and semiconductors[33].

As 1T'-WTe$_2$ has anisotropic lattice structure, polarization resolved transient reflection experiment is initially performed to test the possible anisotropic relaxation dynamics in 1T'-WTe$_2$. The results show that although the amplitude of the signal has polarization angle dependence, the decay dynamics shows no dependence on pump-probe polarization similar to that observed in black phosphorus[34]. Therefore, in the subsequent measurements of the thickness dependent decay dynamics, crystal alignment with respect to the pump-probe polarization is not recorded.

For transient reflection measurement, an infrared optical parametric amplifier (OPA) pumped by a 60 fs, 250 kHz Ti:Sapphire regenerative amplifier, is used in the transient reflection measurement. The idler from OPA at 1940 nm (~150 fs) is used as the probe and the dispersion-compensated residual 800 nm of the OPA is used as the pump. Both pump and probe pulses are linearly polarized. Two half-wave plates are used to alter their polarization angles. A 40X reflective objective lens is used to focus the co-propagating pump and probe spots onto the sample. The reflected probe is collected by the same objective lens and routed through a monochromator followed by an InGaAs photodetector. The detected probe reflection is read by a lock-in amplifier referenced to a 5.7 kHz mechanically chopped pump. The probe spot size is estimated to be 4 μm with a pump spot size slightly larger. The pump excited carrier density is estimated to be around $1.3 \times 10^{15}/cm^{-2}$. It is found that the decay dynamics of pump-excited carriers is independent of the sample orientation with respect to the pump-probe polarization, although the absolute amplitude of the signal is orientation-dependent.

Figure 4f shows transient reflection signal as function of sample thickness. Indeed, three stages of carrier relaxation can be clearly identified for all samples. We find that $\tau_1$, which corresponds to rapid intraband relaxation mainly due to electron-electron and electron-phonon scattering, is quite insensitive to the change of sample thickness. In stark contrast, $\tau_2$ shows marked dependence on sample thickness. This observation indicates the band gap gets less negative as the sample becomes thinner. For very thick sample $\tau_2$ approaches $\tau_1$ indicating that rates of these processes are almost identical. This is consistent with the fact that the bulk material has a negative gap (Figure 4d). As the thickness decreases to 9 nm, the interband recombination process is 4 times slower than thick samples, providing strong indication that the gap becomes less negative and quite likely turns positive in ultrathin 1T'-WT$_2$.

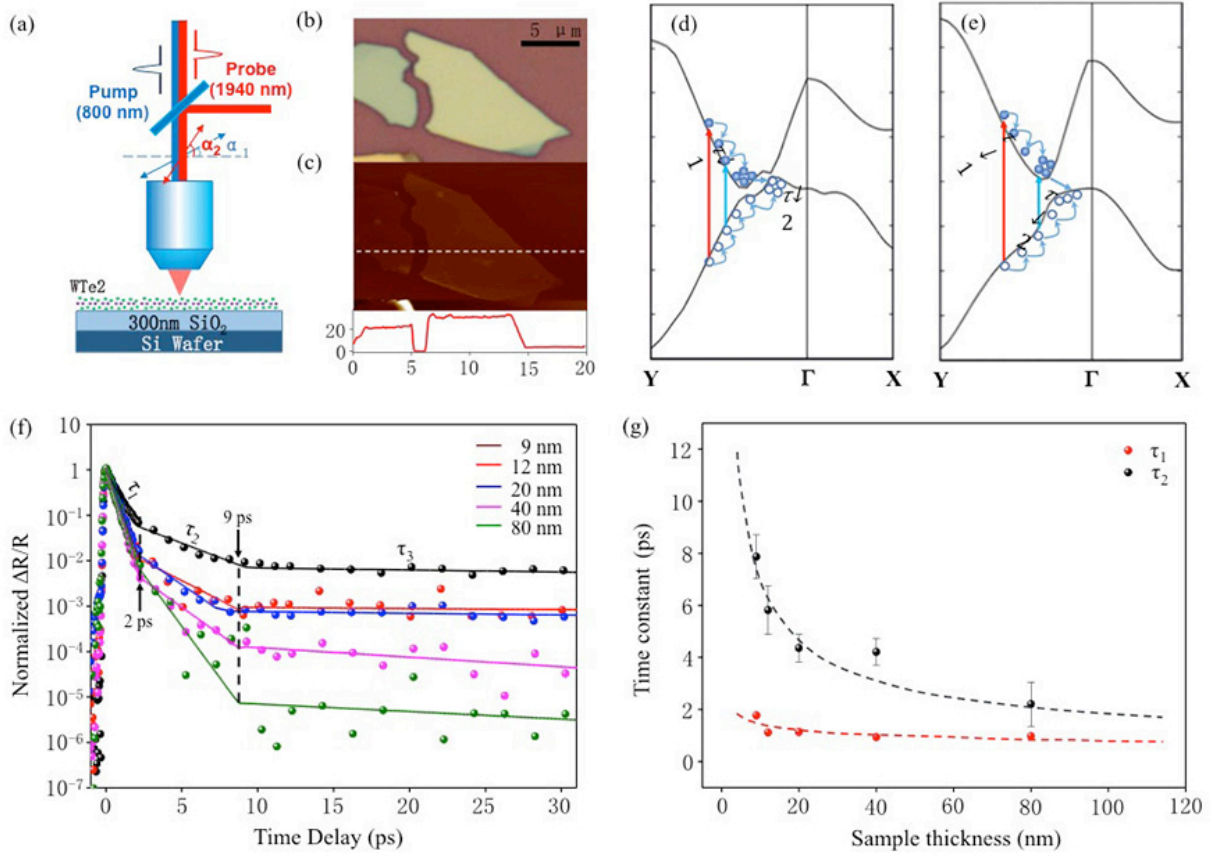

**Figure 4.** Experimental data of optical measurements. (a) Schematic diagram of transient reflection experiment. (b) The optical micrograph and (c) Atomic force micrograph of a sample used with thickness of 20 nm. (d, e) Schematic band diagram and carrier relaxation dynamics of bulk (d) and monolayer (e) 1T'-WTe$_2$ crystal and pump (red) probe (blue) photon transition configuration. (f) The normalized transient reflection spectrum of 1T'-WTe$_2$ with different thicknesses. The data are fitted by exponential decay function at different time span with time constant $\tau_1, \tau_2$ and $\tau_3$. (g) $\tau_1, \tau_2$ as functions of the sample thickness. The dashed curves only serve as guides of eyes.

In the second experiment, we investigate the electrical transport properties of the ultra-thin 1T'-WTe$_2$ samples using standard two-probe configuration. We take extreme care in the preparation of 1T'-WTe$_2$ samples. Ultra-thin 1T'-WTe$_2$ samples are exfoliated from bulk 1T'-WTe$_2$ crystals and transferred to high quality hexagonal boron nitride substrates[35] in a glove box with nitrogen environment, and are covered with PMMA each time the samples needed to be taken out of the glove box briefly. Standard e-beam lithography and the subsequent thermal evaporation were used to define Cr/Au contacts on the samples. The transport measurements were performed in a Quantum Design PPMS system with the sample in dry and low pressure helium atmosphere with the PMMA protection layer to ensure no sample oxidation occurs. The I-V characteristics are collected using a Keithley 2400 source meter. Raman data on similar samples after they have been fabricated into devices and after the transport measurement are provided in Figure 3b, which does not show any measureable degradation of our samples.

Figure 5a shows I-V characteristics of a 11.7-nm thick 1T'-WTe$_2$ sample in temperatures ranging from 1.9 K to 300 K. The insets in Figure 5a are an atomic force microscopy micrograph of the 1T'-WTe$_2$ device with a sample thickness of 11.7 nm. At lower temperatures, I-V curves become patently nonlinear (Figure 5b), which implies non-Ohmic transport. In comparison, another sample with a thickness of 30 nm shows Ohmic behavior at temperatures down to 1.8 K (inset in Figure 5b), consistent with the fact that the bulk material is semimetallic. The low-temperature non-linear I-V curve is typical of formation of a metal-semiconductor junction, or Schottcky contact, between electrode and sample. The 2D thermionic emission equation described by

$$I_{sd} = AA^*_{2D}T^{3/2}\exp\left[-\frac{q}{k_BT}\left(\phi_B - \frac{V_{sd}}{n}\right)\right] \quad (1)$$

where $A$ is the area of the 2D material, $A^*_{2D}$ is the two-dimensional equivalent constant, $k_B$ is Boltzmann constant, $q$ is carrier charge, $n$ is the ideality factor, $V_{sd}$ is source-drain bias, and $\Phi_B$ is the Schottky barrier. To obtain the $\Phi_B$, we plot $\ln(I_{sd}/T^{3/2})$ versus $1000/T$ for different $V_{sd}$, and as shown in Figure 5c, the entire data set can be well fitted linearly. The slopes of linear fits are then plotted against $V_{sd}$ in Figure 5d. According to the equation (1), the intercept at $V_{sd}$ = 0 is equivalent to $-q\Phi_B/1000k_B$, leading to a Schottcky barrier $\Phi_B$ = 1.76 meV for our 11.7-nm thick 1T'-WTe$_2$ sample. Both the 11.7 nm sample and the 30 nm sample are fabricated into devices with identical microfabrication process. The key observation here is the fact that Schottcky barrier forms only in the thinner sample whereas Ohmic contact forms in the thicker sample. This contrast is consistent with the possibility that an energy band gap is opening as 1T'-WTe$_2$ becomes thinner. Although it is impossible to completely rule out the possibility of weak contact with the Cr/Au leads in the measurement, the fact that samples 30 nm or thicker consistently displays Ohmic character when electrodes are fabricated with identical process makes the above scenario highly unlikely.

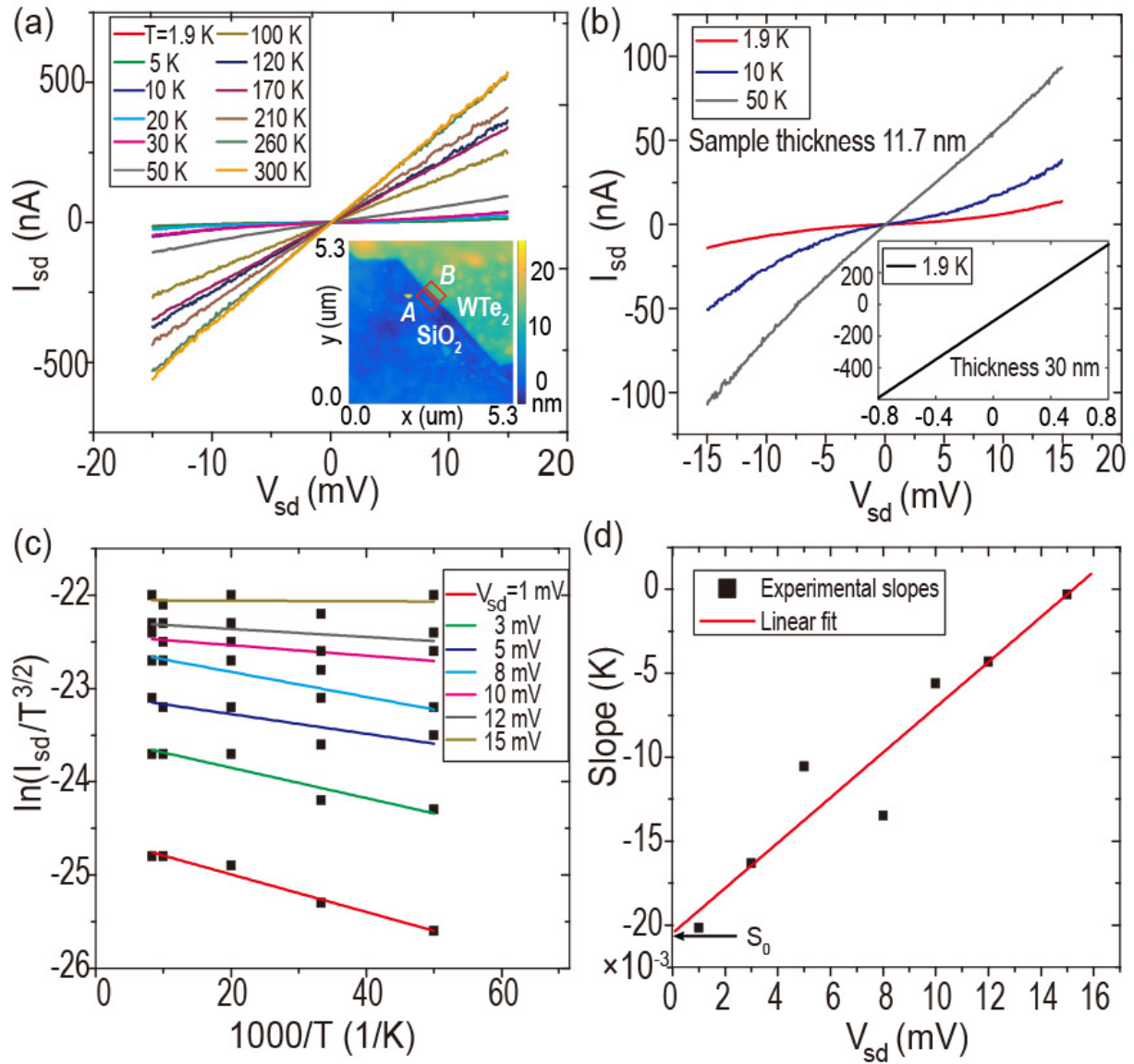

**Figure 5.** Experimental data of electrical transport measurements. (a, b) I-V curves of an 11.7 nm thick 1T'-WTe$_2$ sample, (a) from 1.8 K to 300 K and (b) three low temperatures (1.8 K,

10 K, 50 K). Inset of (a): topography obtained with atomic force microscopy. The thickness of the 1T'-WTe$_2$ sample is obtained by the difference between the averaged hights in region B (1T'-WTe$_2$ sample) and A (SiO$_2$ substrate), enclosed by two red rectangles. Inset of (b): I-V curve of another thicker 1T'-WTe$_2$ sample with thickness of 30 nm. (c) Arrhenius plot of ln($I_{sd}/T^{3/2}$) versus the inverse of temperature and linear fits. (d) Slopes of linear fits in (c), $S_0$ is the extrapolated interception of the slopes at $V_{sd} = 0$, which gives the value of Schottky barrier of the 1T'-WTe$_2$ device.

In summary, our computational and experimental results all point to one very promising possibility, that monolayer 1T'-WTe$_2$ is a quantum spin Hall insulator with a positive band gap. The hybrid functional calculations show that a > 100 meV gap forms in fully relaxed monolayer 1T'-WTe$_2$. Microscopically, this gap is clearly driven by many-body interactions and spin-orbit coupling. The optical and transport measurements are performed on ultrathin (toward monolayer limit) 1T'-WTe$_2$, and provide indirect but consistent evidence for the possible gap opening. The hybrid functional calculations on few-layer 1T'-WTe$_2$ show that the band gap becomes less negative as the layer number decreases. Although the thinnest samples in the reported experiments are still over 10 layers, the optically probed transient carrier dynamics clearly demonstrates that the gap becomes less negative as the sample becomes thinner. The transport measurement confirms the emergence of Schottky barrier in transport measurement in the thinnest sample available.

All these results consistently point to a very attractive possibility that monolayer 1T'-WTe$_2$ becomes fully gapped. Like other transition metal dichalcogenides, 1T'-WTe$_2$ can (though not yet) be isolated into monolayer samples, and thus bears the potential of becoming a truly two-dimensional quantum spin Hall system. In contrast to other transition metal dichalcogenides, WTe$_2$ is the only compound of this family that occurs naturally and stably in the 1T' phase. Combined with the positive band gap, it is clear that monolayer 1T'-WTe$_2$ an attractive for achieving stable free-standing quantum spin Hall phase. An additional remark is in order here, regarding the observation that the band gap of this material is sensitive to in-plane strain[13], but the topology of the band structure is protected by the large gap at Γ, indicating elastic strain engineering can be fruitfully applied to tailor the quantum transport once monolayer 1T'-WTe$_2$ can be isolated. As the bulk 1T'-WTe$_2$ becomes superconducting in the deep compression regime[36, 37], it is also of particular interest then to assay the possibility of inducing non-trivial superconductivity in monolayer 1T'-WTe$_2$ using elastic strain.

**Acknowledgements**

This project has been supported by the National Basic Research Program of China (973 Grant Nos. 2013CB921900, 2012CB921300 and 2014CB920900), the National Natural Science Foundation of China (NSFC Grant Nos. 11174009, 11274015 and 11374021), the Recruitment Program of Global Experts, Beijing Natural Science Foundation (Grant No. 4142024) and the Specialized Research Fund for the Doctoral Program of Higher Education of China (Grant No.20120001110066). K.W. and T.T. acknowledge support from the Elemental Strategy Initiative conducted by the MEXT, Japan and a Grant-in-Aid for Scientific Research on Innovative Areas "Science of Atomic Layers" from JSPS. We wish to thank Kaige Hu, Chaokai Li and Yiwen Wei for useful discussions. The computational work was performed on TianHe-1 (A) at the National Supercomputer Center in Tianjin.

The table of contents entry
**Positive quantum spin Hall gap in monolayer 1T'-WTe$_2$** is consistently supported by density-functional theory calculations, ultrafast pump-probe and electrical transport measurements. It is argued that monolayer 1T'-WTe$_2$, which was predicted to be a semimetallic quantum spin Hall material, is likely a truly 2-dimensional quantum spin Hall insulator with a positive quantum spin Hall gap.

Keyword
qunatum spin Hall effect; density functional theory; ultra-fast spectroscopy; Schottcky junction

*Feipeng Zheng, Chaoyi Cai, Shaofeng Ge, Xuefeng Zhang, Xin Liu, Hong Lu, Yudao Zhang, Jun Qiu, Takashi Taniguchi, Kenji Watanabe, Shuang Jia, Jingshan Qi, Jian-Hao Chen,[*] Dong Sun[†] and Ji Feng[‡]*

E-mail:
[*] chenjianhao@pku.edu.cn;
[†] sundong@pku.edu.cn;
[‡] jfeng11@pku.edu.cn;

**Title**

On the quantum spin Hall gap of monolayer 1T'-WTe$_2$

ToC figure

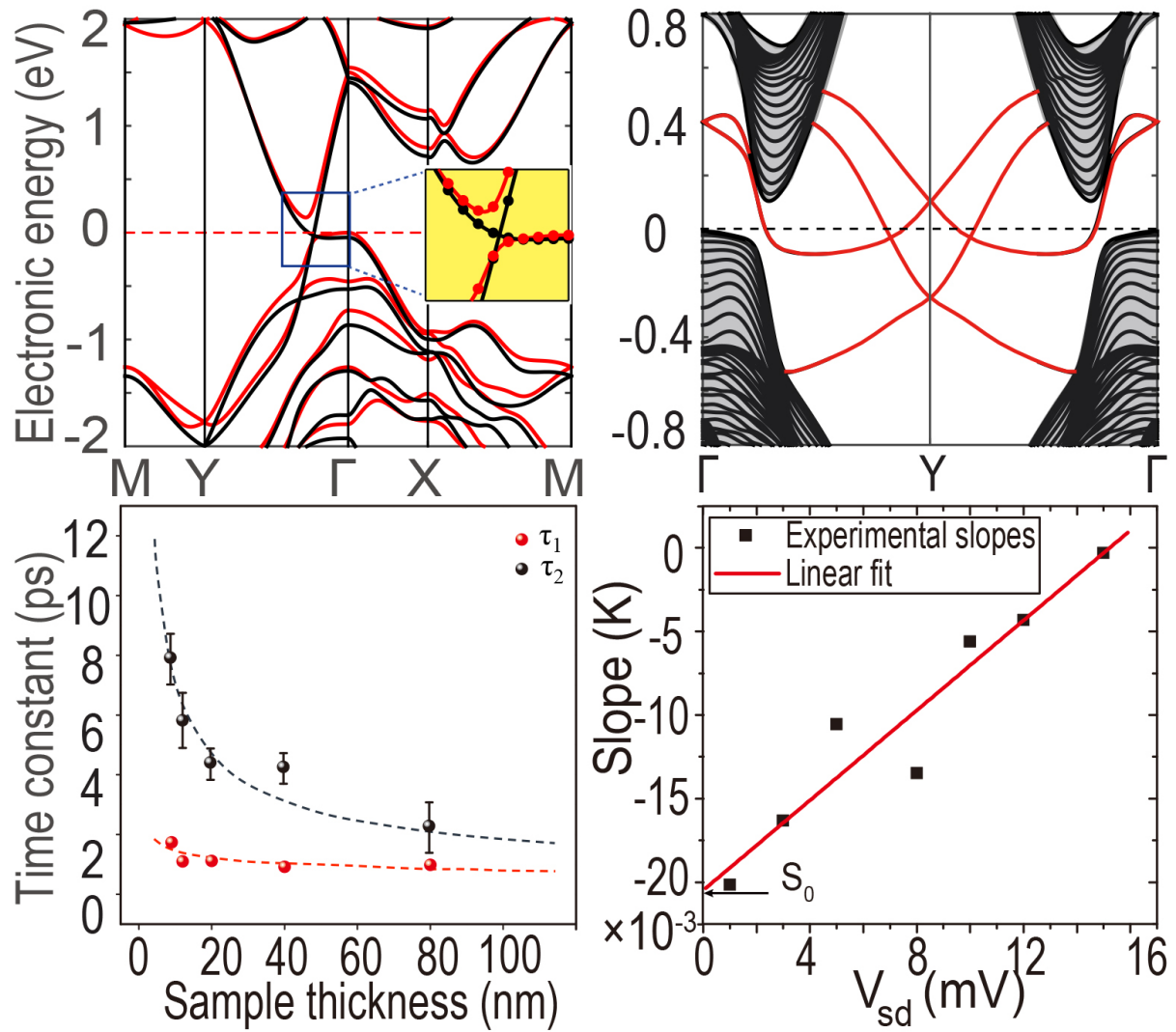